\pdfoutput=1

\documentclass[11pt]{article}
\usepackage{multirow}
\usepackage{tcolorbox}
\usepackage[final]{ACL2023}
\usepackage{amssymb}
\usepackage{times}
\usepackage{latexsym}
\usepackage{xcolor}
\usepackage[utf8]{inputenc}
\usepackage{url}  
\makeatletter
\renewcommand{\@fnsymbol}[1]{\ensuremath{\scriptscriptstyle\ifcase#1\or *\or \dagger\or \ddagger\else\@ctrerr\fi}}  
\makeatother
\usepackage{microtype}

\usepackage{inconsolata}
\usepackage{amsmath}
\usepackage{graphicx}

\usepackage{float}

\usepackage{booktabs}

\title{HASH-RAG: Bridging Deep Hashing with Retriever for Efficient, Fine Retrieval and Augmented Generation}

\author{Jinyu Guo\textsuperscript{1}, \ Xunlei Chen\textsuperscript{1,*}, \ {Qiyang Xia\textsuperscript{1,}}\thanks{~~These authors contributed equally to this work.}, \ Zhaokun Wang\textsuperscript{1},  \ Jie Ou\textsuperscript{1} \\
\textbf{Libo Qin\textsuperscript{2}}, \ \textbf{Shunyu Yao\textsuperscript{3}}, \  \textbf{Wenhong Tian\textsuperscript{1}}\thanks{~~Corresponding author: tian\_wenhong@uestc.edu.cn} \\
  \textsuperscript{1}University of Electronic Science and Technology of China 
  \textsuperscript{2} Central South University  \\ 
  \textsuperscript{3} Big data and artificial intelligent institute, China Telecom Research Institute \\
  \text{guojinyu@uestc.edu.cn}
  }
  
\begin{document}
\maketitle
\begin{abstract}
Retrieval-Augmented Generation (RAG) encounters efficiency challenges when scaling to massive knowledge bases while preserving contextual relevance. We propose Hash-RAG, a framework that integrates deep hashing techniques with systematic optimizations to address these limitations. Our queries directly learn binary hash codes from knowledgebase code, eliminating intermediate feature extraction steps, and significantly reducing storage and computational overhead. Building upon this hash-based efficient retrieval framework, we establish the foundation for fine-grained chunking. Consequently, we design a Prompt-Guided Chunk-to-Context (PGCC) module that leverages retrieved hash-indexed propositions and their original document segments through prompt engineering to enhance the LLM's contextual awareness. Experimental evaluations on NQ, TriviaQA, and HotpotQA datasets demonstrate that our approach achieves a 90\% reduction in retrieval time compared to conventional methods while maintaining considerate recall performance. Additionally, the proposed system outperforms retrieval/non-retrieval baselines by 1.4-4.3\% in EM scores. \thanks{Code available at \url{https://github.com/ratSquealer/HASH-RAG}.}
\end{abstract}

\section{Introduction}

In the era of rapidly expanding data, an increasing number of downstream tasks rely on large language models (LLMs). Within these tasks, Retrieval-Augmented Generation (RAG) is a popular technical framework that incorporates external knowledge sources to tackle knowledge-intensive problems \cite{lewis2020retrieval}. By combining a non-parametric retrieval module with the main model, RAG effectively alleviates hallucination issues in large models \cite{yao2022react, bang2023multitask}. Moreover, this retrieval-and-generation mechanism expands the capabilities of LLMs in few- or zero-shot settings \cite{brown2020language, chowdhery2023palm}, which is now widely considered a standard solution for addressing factual shortcomings in traditional LLMs \cite{ma2023query}.

\begin{figure}[t]
\centering
\includegraphics[scale=0.27]{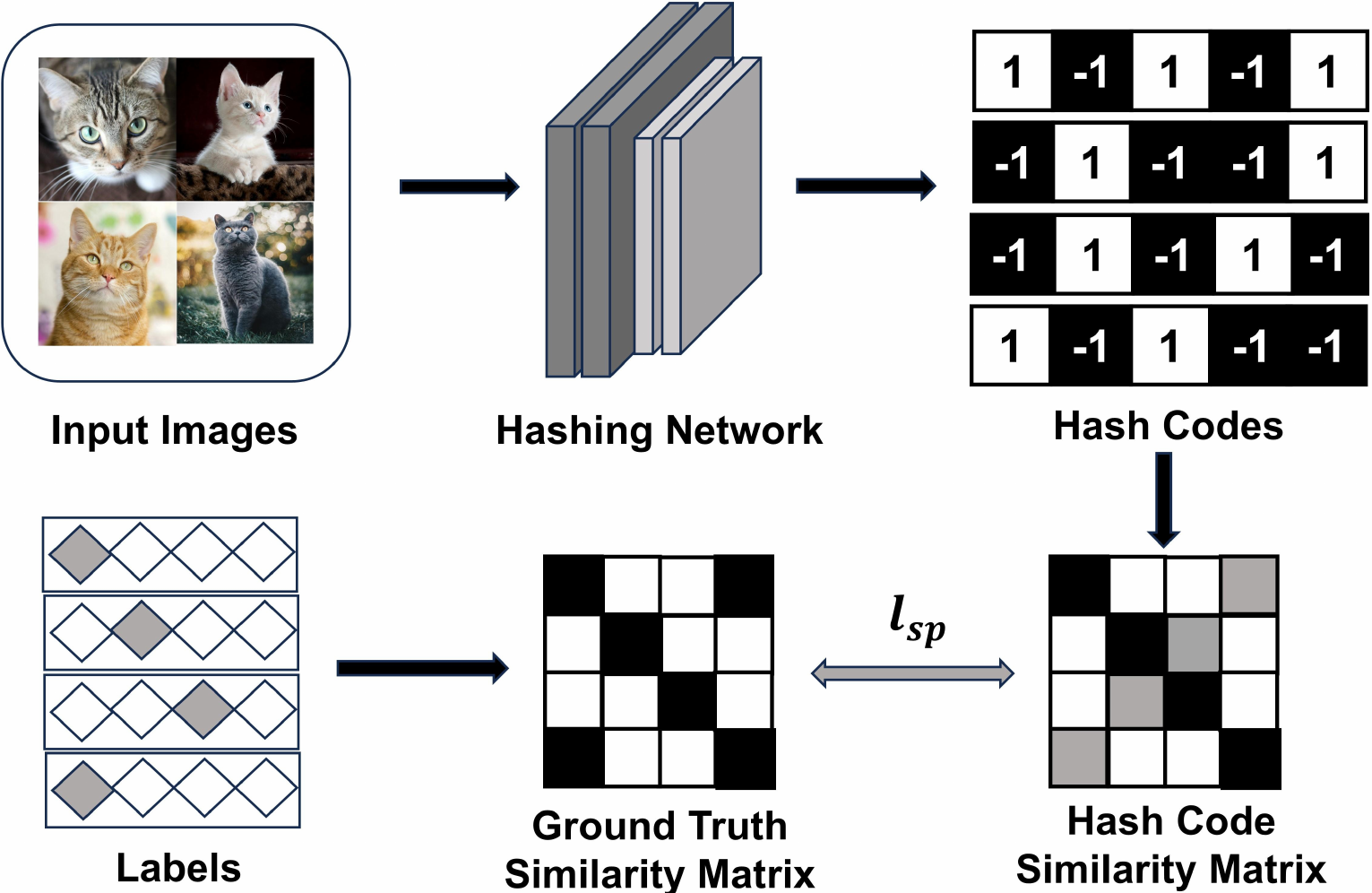}
\caption{Framework of Deep Supervised Hashing with Pairwise Similarity. The framework computes similarity-preserving loss by aligning pairwise relationships between hash codes and their corresponding ground truth, 
which provide a novel perspective for optimizing RAG efficiency.}
\label{Hash_framwork}
\end{figure}

Behind the effectiveness of RAG, the huge scale of the knowledge bases is to ensure the quality and professionalism of the output \cite{izacard2023atlas}. The remarkable effectiveness of RAG systems is fundamentally rooted in the substantial scale of the knowledge bases, which critically ensures output quality and domain expertise. Concurrently, the rapid expansion of both model architectures and knowledge data has established the continuous scaling of RAG knowledge bases \cite{chen2021spann}. Consequently, achieving optimal efficiency-performance coordination in knowledge base retrieval has emerged as a paramount concern in modern RAG system design.

Current research has dedicated significant efforts to optimize the retrieval process in RAG systems. Within them, Several approaches focus on multi-round retrieval iterations to obtain more comprehensive and higher-quality content \cite{yao2023react}. Others employ task-specific or domain-adapted model fine-tuning to enhance performance in targeted scenarios. Alternative strategies involve implementing diversified retrieval techniques or aggregating information from heterogeneous sources \cite{huang2023retrieval}. Chunk optimization strategies demonstrate effectiveness in improving retrieval results by adjusting the chunk size \cite{ sarthi2024raptor} while mitigating risks of model performance degradation \cite{zhao2024retrieval}. Despite these advancements in performance enhancement, RAG systems consistently encounter significant efficiency challenges as knowledge bases expand.

In the domain of large-scale data retrieval, Approximate Nearest Neighbor (ANN) search attracts significant attention due to its capability to substantially reduce search complexity under most scenarios \cite{do2016learning, wang2017survey}. Among ANN techniques \cite{luo2021deep}, hashing methods emerged as one of the most widely adopted approaches, offering exceptional storage efficiency and rapid retrieval capabilities \cite{cantini2021learning}. In particular, deep hashing methods \cite{xia2014supervised} employ deep neural networks to learn discriminative feature representations and convert them into compact hash codes, substantially enhancing retrieval efficiency while reducing storage requirements and computational costs \cite{lai2015simultaneous}. Figure \ref{Hash_framwork} illustrates a typical deep supervised hashing framework. This approach marks an unprecedented breakthrough in large-scale image retrieval and significantly outperforms conventional method \cite{chen2021spann}, which also provide a novel perspective for optimizing RAG efficiency.

In the era of RAG, the knowledge bases significantly surpass traditional image retrieval datasets in scale and growth velocity. In light of this, ANN techniques exemplified by hashing demonstrate significant potential for RAG applications with their capacity to rapidly target results and reduce computational complexity in large-scale data processing.

In this paper, we introduce ANN-based techniques into RAG frameworks and propose Hash-RAG through systematic integration of deep hashing methods. Specifically, our architecture converts query embeddings into binary hash codes via sign function operations. For knowledge bases, we adopt an asymmetric processing strategy to optimize training efficiency by directly learning binary hash codes without feature learning. Based on this, we achieve fine-grained retrieval through corpus chunking, which filters redundant content while preserving precision. Nevertheless, we notice that existing chunking approaches result in retrieved segments lacking essential contextual information, which substantially degrades the quality of generated outputs. To address this, we also propose a Prompt-Guided Chunk-to-Context (PGCC) module, which splits documents into factual fragments (i.e., propositions) as retrieval units. These propositions are structured in a concise, self-contained natural language format and indexed to their original documents. During generation, LLM processes hash-based retrieved propositions and their contexts through specifically designed prompts to generate, achieving optimal coordination between accuracy and efficiency.

We conducted experiments on open-domain question-answering datasets, including Natural Questions (NQ) \cite{kwiatkowski2019natural}, TRIVIAQA \cite{joshi2017triviaqa}, and the more complex multi-hop HOTPOTQA \cite{yang2018hotpotqa}.
Experimental results show that our model significantly reduces retrieval time, which requires only 10\% of the time needed for conventional retrieval methods while maintaining advanced recall performance.
In combination with the PGCC module, we have achieved a performance increase in the generation task while retaining efficiency.

Main contributions of this paper are as follows:

\begin{enumerate}
    \item We propose HASH-RAG, a framework that systematically integrates deep hashing into RAG with stage-wise optimizations. 
    This approach significantly enhances computational efficiency in large-scale knowledge retrieval, thereby accelerating end-to-end inference throughout the overall RAG.
    \item Building upon our hash-based efficient retrieval framework, we propose the PGCC module that enables fine-grained retrieval while enhancing contextual information through prompt-based optimization.
    \item Experimental results on multiple datasets demonstrate that HASH-RAG significantly improves the efficiency of the retrieval. With PGCC module, our method surpasses RAG baseline models in overall performance, achieving optimal coordination between efficiency and performance.
\end{enumerate}

\section{Related Work}
\subsection{Retrieval-Augmented Generation}

RAG mitigates LLM hallucinations through non-parametric memory integration and compensates for factual deficiencies via external knowledge retrieval \cite{gao2023retrieval}. Early implementations relied on statistical similarity metrics (TF-IDF \cite{robertson1997relevance}, BM25 \cite{robertson2009probabilistic}) before transitioning to vectorized representations \cite{karpukhin2020dense}, enabling end-to-end tunable systems with context-aware retrieval. Recent efforts focus on two phases: pre-retrieval query-data matching to enhance precision \cite{ma2023query} and post-retrieval content re-ranking or transformation to optimize generator input \cite{glass2022re2g}. A persistent challenge lies in chunking strategy design, balancing granularity trade-offs: coarse chunks risk redundancy despite contextual richness \cite{shi2023large}, while fine-grained units sacrifice semantic completeness for precision \cite{raina2024question}. Scalability-induced efficiency bottlenecks in retrieval optimization now critically constrain RAG advancement.

\subsection{Deep Hashing Methods}

Approximate Nearest Neighbor (ANN) algorithms address large-scale search inefficiency by trading exact precision for efficiency. Unlike tree-based (Annoy \cite{bernhardsson2015annoy}), quantization-based (PQ \cite{jegou2010product}), or graph-based (HNSW \cite{malkov2018efficient}) methods, hashing reduces memory requirements and search latency while preserving vector locality, which makes it a mainstream ANN solution. Early techniques like Locality-Sensitive Hashing (LSH) \cite{charikar2002similarity, indyk1998approximate} relied on predefined mappings to hash buckets, requiring multiple tables for satisfactory recall. Learning-to-hash methods (e.g., Spectral Hashing \cite{weiss2008spectral}, Semantic Hashing \cite{salakhutdinov2009semantic}) later optimized hash functions to improve retrieval efficiency and accuracy. Current research focuses on deep-supervised hashing methods 
(e.g., Convolutional neural network hashing (CNNH) \cite{xia2014supervised}: two-phase of binary codes generation and convolutional neural networks (CNNs) training, Deep supervised hashing (DSH) \cite{liu2016deep}: pairwise similarity loss with regularization for end-to-end training, Maximum-margin hamming hashing (MMHH) \cite{kang2019maximum}: discriminative enhancement through t-distribution and semi-batch optimization). These methods demonstrate adaptability to large-scale retrieval, establishing technical foundations for optimizing RAG’s chunking strategies and retrieval efficiency.

\begin{figure*}[ht]
\centering
\includegraphics[scale=0.47]{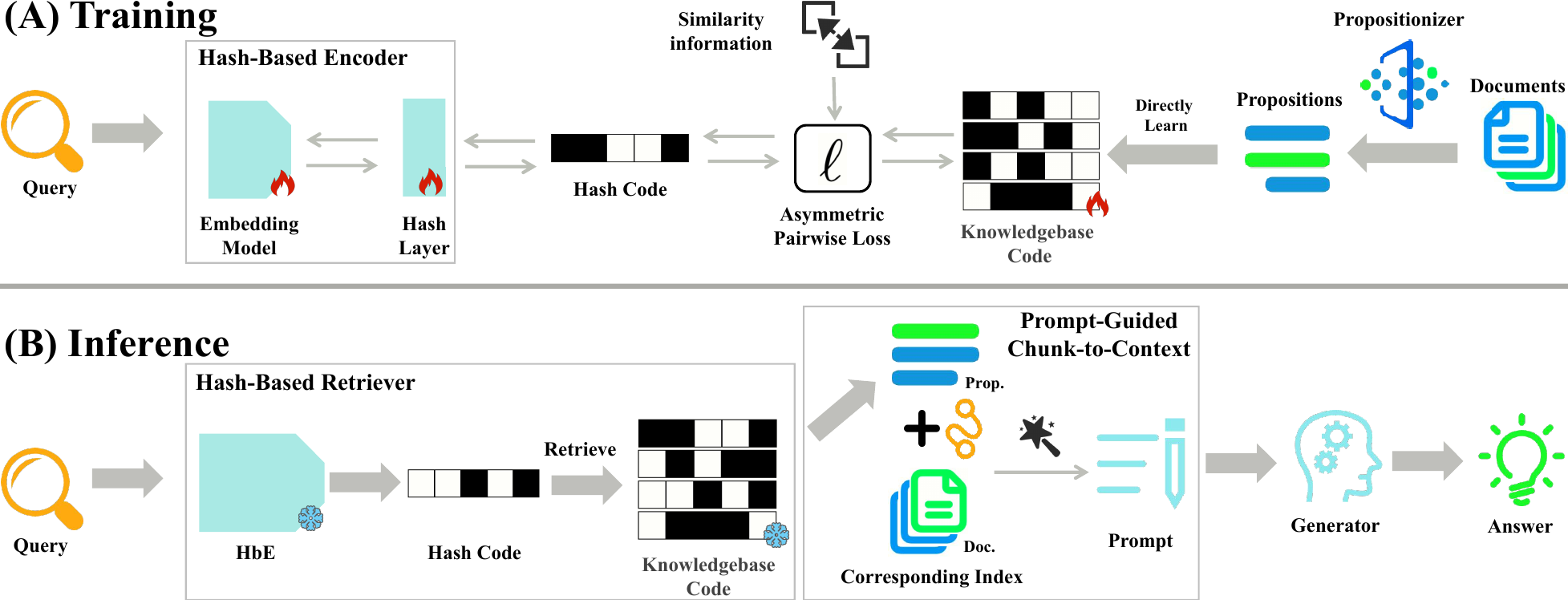}
    \caption{Framework Overview. (a) Training. The hash-based encoder generates compact query hash codes, while the knowledge base creates binarized propositional codes from factually chunked corpora. Both components are jointly optimized through an asymmetric pairwise loss with similarity constraints. (b) Inference. The hash-based retriever efficiently fetches relevant propositions, augmented by indexed document references for contextual grounding. The generator synthesizes evidence from these elements using optimized prompts to produce responses.}
\label{pipeline}
\end{figure*}

\section{Method}

In this study, we propose a Hash Retrieval-Augmented Generation (Hash-RAG) framework, a pipeline that enhances RAG from the perspectives of deep hashing and contextual enhancement. Figure \ref{pipeline} illustrates the comprehensive architecture of Hash-RAG. Sections 3.1 and 3.2 respectively introduce the Hash-Based Retriever (HbR) and the module Prompt-Guided Chunk-to-Context (PGCC).

\subsection{Hash-Based Retriever}
\noindent \textbf{Hash-based Encoder (HbE)}\ \ \ HbE module comprises two components: a query encoder $E_q$ and a proposition encoder $E_p$, which generate binary hash codes for queries $q$ and propositions $p$ (derived from knowledge base document segmentation), respectively. Inspired by ADSH \cite{jiang2018asymmetric}, we design asymmetric encoding strategies for queries and propositions. The query encoder $E_q$ employs BERT-base-uncased \cite{kenton2019bert} as its embedding model to map queries into a $d$-dimensional vector space ($d = 768$). The query embedding vector $v_q \in \mathbb{R}^d$:

\begin{equation}
v_q = BERT(q, \theta)
\end{equation}

\noindent where $\theta$ denotes BERT's parameter. Subsequently, the binary hash code of query $q$ is computed through the sign function in the hashing layer: $h_q = sign(v_q)$. To address the vanishing gradient problem of the sign function, we approximate it using a scaled $tanh$ function \cite{cao2017hashnet}. 

\begin{equation}
\widetilde{h_q}=tanh(\beta {v_q}) \in \{-1, 1\}^l
\end{equation}

\noindent where $l$ is the fixed hash code length (768 bits) and $\beta$ is the scaling hyperparameter controlling approximation smoothness ($\beta \to \infty$ converges to the sign function). We set $\sigma = 0.1$ and $\beta = \sqrt{\sigma \cdot step + 1}$, where $step$ counts completed training steps.

The proposition encoder $E_p$ directly learns binary hash codes without embedding model training through specialized loss functions and an alternating optimization strategy to reduce training overhead. For training data $\Delta = \{ \langle q_i, p_i^+, \{p_{i,j}^-\}_{j=1}^n \rangle \}_{i=1}^m$ containing $m$ instances, the supervision matrix $S \in \{-1,1\}^{m \times (n+1)}$ labels positive ($S_{i,j}=1$) and negative ($S_{i,j}=-1$) samples. We minimize $L_2$ loss between binary code inner products and supervision information:

\begin{equation}
\mathcal{L}_{HbE} = \sum_{i = 1}^{m} \sum_{j = 1}^{n} \left[ {\widetilde{h_{q_i}}}^T h_{p_j} - lS_{ij} \right]^2
\end{equation}

\noindent where $h_p$ denotes the proposition hash code. For proposition-only training data $\Delta_p = \{q_j\}_{j = 1}^n$, we construct simulated queries by randomly sampling $m$ propositions from all proposition indices set $\Gamma = \{1, 2, \ldots, n\}$, forming index subset $\Omega = \{i_1, i_2, \ldots, i_m\} \subseteq \Gamma$. The extended loss function is as follows:  

\begin{equation}
\begin{aligned}
\mathcal{L}_{HbE} &= \sum_{i\in\Omega} \sum_{j\in\Gamma} \left[ \tanh(\beta v_{p_i})^T h_{p_j} - lS_{ij} \right]^2 \\
&+ \gamma \sum_{i\in\Omega} \left[ h_{p_j} - \tanh(\beta v_{p_i}) \right]^2
\end{aligned}
\label{formula:L_Hbe2}
\end{equation}

\noindent where $\gamma$ constrains $h_{p_i}$ and $\widetilde{h_{p_i}} = \tanh(\beta v_{p_i})$ to be as close as possible, which enables effective optimization of proposition hash codes through iterative parameter updates.

We implement an alternating optimization strategy, alternately fixing and updating the neural network parameters $\theta$ and the proposition sentence hash code matrix $\boldsymbol{H}$.

\noindent \textbf{Update $\theta$ with $\boldsymbol{H}$ fixed}\ \ \ With $\boldsymbol{H}$ fixed, we compute the gradient of $\mathcal{L}_{HbE}$ w.r.t. $v_{p_i}$:  

\begin{equation}
\begin{aligned}
\frac{\partial \mathcal{L}_{HbE}}{\partial v_{p_i}} &= \Biggr\{2 \sum_{j\in\Gamma} \left[ \left( {\widetilde{h_{p_i}}}^T h_{p_j} - lS_{ij} \right) h_{p_j} \right] \Biggr. \\
&\Biggr. + 2\gamma \left( \widetilde{h_{p_i}} - h_{p_i} \right) \Biggr\} \odot \left( 1 - {\widetilde{h_{p_i}}}^2 \right)
\end{aligned}
\end{equation}

The chain rule propagates this gradient through BERT's parameters $\theta$, which are then updated via backpropagation.

\noindent \textbf{Update $\boldsymbol{H}$ with $\theta$ fixed}\ \ \ With $\theta$ fixed, we rewrite Equation \ref{formula:L_Hbe2} in matrix form:

\begin{equation}
\begin{aligned}
\mathcal{L}_{HbE} &= \|\widetilde{\boldsymbol{V}}\boldsymbol{H}^T - l\boldsymbol{S}\|_F^2 + \gamma\|\boldsymbol{H}^{\Omega} - \widetilde{\boldsymbol{V}}\|_F^2 \\
&= \|\widetilde{\boldsymbol{V}}\boldsymbol{H}^T\|_F^2 - 2l\mathrm{tr}(\boldsymbol{H}^T\boldsymbol{S}^T\widetilde{\boldsymbol{V}}) \\
&-2\gamma\mathrm{tr}(\boldsymbol{H}^{\Omega}\widetilde{\boldsymbol{V}}^T) + \mathrm{const}
\end{aligned}
\end{equation}

\noindent where $\widetilde{\boldsymbol{V}} = \left[ \widetilde{v_{p_1}}, \widetilde{v_{p_2}}, \ldots, \widetilde{v_{p_m}} \right]^T \in [-1, +1]^{m \times l}$, and $\boldsymbol{H}^{\Omega} = \left[ h_{p_1}, h_{p_2}, \ldots, h_{p_m} \right]^T$ denotes the sampled proposition hash codes.

To update $\boldsymbol{H}$, we adopt a column-wise updating strategy. For the $k$-th column $\boldsymbol{H}_{*k}$ with residual matrices $\widehat{\boldsymbol{H}}_k$ (excluding column $k$), and the $k$-th column $\widetilde{\boldsymbol{V}_{*k}}$ with residual matrices $\widehat{\boldsymbol{V}}_k$ (excluding column $k$), the optimisation objective function is:

\begin{equation}
\begin{aligned}
\mathcal{L}(\boldsymbol{H}_{*k}) = &\mathrm{tr}\left( \boldsymbol{H}_{*k} \left[ 2\widetilde{\boldsymbol{V}_{*k}}^T \widehat{\boldsymbol{V}}_{k} \widehat{\boldsymbol{H}_{k}}^T + \boldsymbol{Q}_{*k}^T \right] \right) \\
& + \mathrm{const}
\end{aligned}
\end{equation}

\noindent where $\boldsymbol{Q} = -2l\boldsymbol{S}^T\widetilde{\boldsymbol{V}} - 2\gamma\overline{\boldsymbol{V}}$, with the $k$-th column $\boldsymbol{Q}_{*k}$. The optimal solution is:

\begin{equation}
\begin{aligned}
\boldsymbol{H}_{*k} = - \mathrm{sign}\left( 2\widehat{\boldsymbol{H}}_{k} \widehat{\boldsymbol{V}_{k}}^T \widetilde{\boldsymbol{V}_{*k}} + \boldsymbol{Q}_{*k} \right)
\end{aligned}
\end{equation}

The alternating optimization between $\theta$ and $\boldsymbol{H}$ drives gradual convergence through multiple iterations, ultimately producing an effective query hash function and robust proposition hash code.

\subsection{Prompt-Guided Chunk-to-Context}

\noindent \textbf{Retrieval Unit Granularity}\ \ \ We employ the information bottleneck theory \cite{tishby2000information} to optimize retrieval unit selection, where proposition-based chunks preserve maximal generator-relevant information while minimizing noise. Given the joint probability distribution $p(X,Y)$ between document $X$ and generator output $Y$, we quantify the information content about  $Y$ contained within compressed proposition \( \widetilde{X} \) through mutual information:

\begin{equation}
I(\widetilde{X};Y)=\int_{\widetilde{\mathcal{X}}}\int_{\mathcal{Y}}p(\widetilde{x},y)\log\frac{p(\widetilde{x},y)}{p(\widetilde{x})p(y)}d\widetilde{x}dy
\end{equation}

The compression objective minimizes $\mathcal{L}_{\mathrm{IB}} = I(\widetilde{X};X) - \beta I(\widetilde{X};Y)$, where the Lagrange multiplier $\beta$ balances information retention and compression. Unlike conventional sentence/paragraph units \cite{karpukhin2020dense}, we adopt proposition units \cite{min2023factscore} that capture atomic semantic expressions. Proposition extraction occurs during the knowledge base preprocessing phase (i.e., index construction). So for document $Doc$, we extract $k$ interrelated propositions $X = [x_1, \ldots, x_k]$, with relevance scores computed through hybrid scoring:

\begin{equation} 
X_f = \alpha X_{doc} + (1 - \alpha) \sum_{k=1}^{n} w_k x_k
\label{eq:scoring_function} 
\end{equation}

\noindent where $X_{doc}$ and $x_k$ denote document-level and BERT-based proposition scores respectively, with $\alpha$ and $w_k$ optimized via cross-validation. 

Hash-based retrieval optimizes proposition selection through Hamming distance relationships:

\begin{equation}
\text{dist}_H(h_{q_i}, h_{p_j}) = \frac{1}{2}(d - \langle h_{q_i}, h_{p_j}\rangle) \label{eq:hamming_distance}
\end{equation}

\noindent where $d$ is the binary code dimension. We iteratively expand the Hamming radius until selecting the top $\alpha$ propositions, $\alpha$ denotes the approximate candidate set retrieved via Hamming distance calculations on hash codes:

\begin{equation}
\text{Top } P_j = \arg\max_{i\in\{1,\ldots,\alpha\}} \langle v_{q_i}, h_{p_j} \rangle
\end{equation}

Deduplication over proposition-document mappings yields the final top $k$ retrieved documents $\{Doc_1, \ldots, Doc_k\} = \text{Duplicates}(P_1 \cup \ldots \cup P_j)$. This dual optimization of semantic compression and hash-based retrieval ensures maximal information extraction with minimal noise.

\noindent \textbf{Prompt Optimization}\ \ \ We employ LLAMA2 as the generator with optimized prompts. The Hash-Retriever identifies top-$j$ propositions $P_j = \{P_1, \ldots, P_j\}$ and their corresponding document indices ${Doc}_k$, forming the generator's context through three key components: (1) \textbf{Additional Prompt} instructing semantic integration of propositions and indexed documents for precise responses, (2) \textbf{Retrieved Segments} containing similarity-ranked propositions $P_j$ with document references ${Doc}_j$, and (3) \textbf{Indexed Documents} ${Doc}_1, \ldots, Doc_k$ providing contextual grounding. 

The prompt template activates the generator's capability through chunk-to-context: propositions supply direct evidence while documents offer broader context, enabling accurate intent understanding with balanced retrieval-context integration. For details, see appendix \ref{Prompt Template}.

\section{Experiment}
\subsection{Experimental Settings}

\begin{table*}[t]
\centering
\begin{tabular}{l|ccc|ccc|ccc|cc}
\toprule
\multirow{2}{*}{\textbf{Model}} & \multicolumn{3}{c|}{\textbf{Top 5}} & \multicolumn{3}{c|}{\textbf{Top 20}} & \multicolumn{3}{c|}{\textbf{Top 100}} & \textbf{Index} & \textbf{Query} \\
  & NQ & TQA & HQA & NQ & TQA & HQA & NQ & TQA & HQA & \textbf{size} & \textbf{time}\\
\midrule
\textbf{BM25} & 45.2 & 55.7 & - & 59.1 & 66.9 & - & 73.7 & 76.7 & - & 7.4  & 913.8 \\
\textbf{SimCSE} & 28.8 & 44.9 & 26.7 & 44.3 & 59.4 & 44.1 & 47.0 & 62.4 & 46.1 & 64.6 & 548.2 \\
\textbf{Contriever} & 47.8 & 59.4 & 42.5 & 67.8 & 74.2 & 67.4 & 82.1 & 83.2 & 76.9 & 64.6 & 608.0 \\
\textbf{PQ} & 50.7 & 63.0 & 43.7 & 72.2 & 73.2 & 64.3 & 81.2 & 80.4 & 62.1 & 2.0 & 46.2 \\
\textbf{MEVI\textsuperscript{$\dagger$}} & \textbf{75.5} & - & - & \textbf{82.8} & - & - & \underline{87.3} & - & - & 151.0 & 222.5 \\
\midrule
\textbf{DPR} & 66.0 & 71.6 & 54.4 & 78.4 & 79.4 & \underline{73.0} & 85.4 & \underline{85.0} & \underline{80.3} & 64.6 & 456.9 \\
\textbf{LSH\textsuperscript{$\ddagger$}} & 43.2 & 48.0 & 38.4 & 63.9 & 65.2 & 60.5 & 77.2 & 76.9 & 71.1 & \textbf{2.0} & \textbf{28.8} \\
\textbf{DSH\textsuperscript{$\ddagger$}} & 57.2 & 64.7 & 44.2 & 77.9 & \underline{77.9} & 66.2 & 85.7 & 84.5 & 80.4 & \underline{2.2} & \underline{38.1} \\
\midrule
\multirow{2}{*}{\textbf{HbR(Ours)}} & 72.4 &  \underline{78.3} & \underline{57.7} & \underline{80.3} & \textbf{87.0} & \textbf{80.2} & \textbf{87.5} & \textbf{88.4} & \textbf{81.4} & \multirow{2}{*}{4.6} & \multirow{2}{*}{42.3} \\
  & $_{\pm0.2}$ & $_{\pm0.2}$ & $_{\pm0.1}$ & $_{\pm0.2}$ & $_{\pm0.1}$ & $_{\pm0.1}$ & $_{\pm0.3}$ & $_{\pm0.1}$ & $_{\pm0.1}$ &  & \\
\bottomrule
\end{tabular}
\caption{Top $k$ recall on test sets with the index size(GB) and query time(ms) of HbR and baselines. $\dagger$ Model selected is MEVI Top-100 \& HNSW from the main experiments and HNSW backbone retrieval method results in a multiplicative increase in index size, and the retrieval time of MEVI is not SOTA. $\ddagger$ Integration of hash with the encoder, selecting DPR, DSH model selected is hash table lookup with candidate = 1000.}
\label{tab:model_performance}
\end{table*}

\begin{table*}[ht]
\centering
\begin{tabular}{lcccccc}
\toprule
\multirow{2}{*}{\textbf{Model}} & \multicolumn{3}{c}{\textbf{LLAMA2-7B}} & \multicolumn{3}{c}{\textbf{LLAMA2-13B}}   \\
  & NQ & TQA & HQA & NQ & TQA & HQA \\
\midrule
$\textbf{ToolFormer}^\lozenge$ & 17.7 & 48.8 & 14.5 & 22.1 & 51.7 & 19.2  \\
\textbf{RRR} & 25.2 & 54.9 & 19.8 & 27.1 & 59.7 & 24.4  \\
\textbf{FILCO} & 25.8 & 55.0 & 19.4 & 27.3 & 60.4 & 23.9  \\
$\textbf{REPLUG}^{\blacklozenge}$ & 27.1 & \underline{57.1} & 20.5 & 29.4 & 62.7 & 26.8 \\
\textbf{Hash-RAG} & \textbf{28.5$_{\pm0.1}$} & \underline{57.1$_{\pm0.1}$} & \textbf{22.1$_{\pm0.2}$} & \textbf{34.9$_{\pm0.2}$} & \textbf{64.5$_{\pm0.1}$} & \textbf{31.1$_{\pm0.3}$} \\
\bottomrule
\end{tabular}
\caption{EM of open-domain QA. $\lozenge$ Generation models in this experiment involve the GPT series, all of which are modified to the LLAMA2 series and $w/o$ train reader in this experiment. $\blacklozenge$ Contriever and a zero-shot setting are selected.}
\label{tab:Gmodel_performance}
\end{table*}

\noindent \textbf{Datasets and Retrieval Corpus}\ \ \ We evaluated our model on three QA benchmarks using development sets. These datasets contain Wikipedia and web-sourced questions, representing diverse knowledge-intensive tasks: NQ \cite{kwiatkowski2019natural} and TRIVIAQA \cite{joshi2017triviaqa} assess direct knowledge recall, while HOTPOTQA \cite{yang2018hotpotqa} requires multi-hop reasoning across documents. Different retrieval granularities from sentences to full documents refer to Appendix \ref{Dataset}.

\noindent \textbf{Metrics}\ \ \  With more retrieval units, we retrieve additional propositions, map them to source documents, deduplicate, and return the top $k$ unique documents. We evaluate using document recall@k and retrieval efficiency (index size/query time). For generation, Exact Match (EM) assesses whether the ground truth appears exactly in the output.

\noindent \textbf{Implementation Details}\ \ \  In our paper, the encoders (Embedding) utilize BERT base, large, ALBERT, and ALBERT, with each model initialized using the official pre-trained weights. The number of top $j$ propositions is fixed at 100. The generator (generator) LLMs include LLaMA2-7B and 13B \cite{touvron2023llama}. For the HbR model used in our primary experiments, the training batch size is set to 128, with one additional BM25 negative passage per question. Each encoder is trained for 40 epochs, employing linear scheduling with warm-up and a dropout rate of 0.1. 

\noindent \textbf{Baselines}\ \ \ We compare HbR with BM25 \cite{robertson2009probabilistic}, DPR \cite{karpukhin2020dense}, SimCSE \cite{gao2021simcse}, Contriever \cite{izacard2021contriever}, Model-enhanced Vector Index (MEVI) \cite{zhang2024model}, LSH \cite{charikar2002similarity}, and DSH \cite{liu2016deep}. 

BM25 \cite{robertson2009probabilistic} employs TF-IDF principles for document relevance ranking, while DPR \cite{karpukhin2020dense} utilizes a dual-encoder architecture; SimCSE \cite{gao2021simcse}, an unsupervised learning architecture, optimizes semantic representations via positive/negative pair discrimination; Contriever \cite{izacard2021contriever} employs Transformer-based encoder and optimize a contrastive loss function; MEVI \cite{zhang2024model} employs the clusters of Residual Quantization (RQ) to search for documents semantically; LSH's \cite{charikar2002similarity} hash-bucket mapping reduces the scope of nearest neighbor search; DSH \cite{liu2016deep} integrates deep feature extraction with semantic label optimization in hash space.

\subsection{Main Result}
Table \ref{tab:model_performance} illustrates HbR's recall@k ($k \in {5,20,100}$) and latency on NQ, TQA, and HotpotQA benchmarks. Our framework reduces query latency to one-tenth of conventional retrievers while achieving 0.2-8.6\% higher recall@20/100. Notably, HbR achieves optimal performance at $k=20$. While H-RAG underperforms MEVI at $k=5$, such small-$k$ scenarios are secondary since users typically require $\geq$20 passages for answer generation. The hashing mechanism maintains index size advantages despite data volume increases from chunking.

Next, Table \ref{tab:Gmodel_performance} compares Hash-RAG with baseline RAG systems using LLAMA2-7B/13B. Our framework outperforms retrieval-optimized methods (FILCO \cite{wang2023learning}, RRR \cite{ma2023query}) and LLAMA-Retrieval hybrids (REPLUG \cite{shi2024replug}), with all retrieval-augmented models surpassing Toolformer's non-retrieval baseline \cite{schick2023toolformer}. By feeding top-20 results to generators, we optimally balance context volume and generation quality. This design choice leverages the complementary strengths of hashing-based retrieval and modern LLMs, demonstrating significant EM improvements across all benchmarks.

\subsection{Ablations}

\noindent \textbf{Encoder Version}\ \ \ We investigated the compatibility of various encoder versions (ALBERT, Bert-base, Bert-large, RoBERTa) with our model. As shown in Table \ref{table:Bert_version}, proposition-level chunking achieves significantly superior retrieval performance compared to sentence-level and paragraph-level strategies in terms of Recall@20 metrics. Although BERT-large achieved the highest Top-k recall rates, to ensure a fair comparison with existing models that employ BERT-base as their encoder architecture, we adopt the identical configuration in our implementation.

\begin{table}[t]
\centering
\begin{tabular}{c|c|c|c}
\hline
\textbf{Model} & Top 5 & Top 20 & Top 100 \\
\hline 
\textbf{ALBERT}       & 63.2 & 78.1 & 82.4 \\
\textbf{Bert-base}  & 72.4 & 80.3 & 87.5 \\
\textbf{RoBERTa}     & 72.6 & 84.7 & 87.6 \\
\textbf{Bert-large} & \textbf{73.1} & \textbf{85.8} & \textbf{87.9} \\
\hline
\end{tabular}
\caption{Top $k$ recall of HbR using different versions of embedding models on NQ Dataset.}
\label{table:Bert_version}
\end{table}

\begin{table}[t]
\centering
\begin{tabular}{l|c}
\hline
Chunk Strategy & Recall@20 \\
\hline 
sentence & 62.9 \\
paragraph & 68.8 \\
prop. & \textbf{80.2}\\
\hline 
\hline
Prompt Optimization & EM \\
\hline 
HbR $w/o$ prop. &25.3 \\
HbR $w/o$ doc. & 24.8 \\
HbR & 29.4\\
HbR $w/$ prompt (Ours) & \textbf{31.1} \\
\hline
\end{tabular}
\caption{Metrics (Recall@20 and EM) of different chunk strategies and prompt optimizations on HotpotQA dataset, with prop. denoting the propositions and doc. representing the original source documents associated with the retrieved propositions.}
\label{table:Chunk_Strategy&Prompt_Optimization}
\end{table}

\noindent \textbf{Chunk Strategy}\ \ \
The performance hierarchy in Table \ref{table:Chunk_Strategy&Prompt_Optimization} demonstrates proposition-level chunking surpassing sentence- and paragraph-level strategies on Recall@20 metrics respectively. Experimental analysis reveals sentence-level segmentation fractures predicate-argument coherence, while paragraph-level processing incorporates extraneous content. The performance hierarchy reflects proposition-level chunking's dual advantage: preserving self-contained semantic units and systematically eliminating contextual noise.

\noindent \textbf{Prompt Optimization}\ \ \
Table \ref{table:Chunk_Strategy&Prompt_Optimization} presents our comparative analysis of prompt optimization strategies, where the PGCC module empirically demonstrates superior EM performance over all baselines. Notably, the $w/o$ doc. configuration outperforms $w/o$ prop., suggesting that in multi-hop datasets under Recall@20 settings, self-contained propositions still require cross-verification with source documents even when guided by optimized prompts. The non-prompted configuration achieves secondary performance due to sufficient contextual data supporting LLM reasoning, while prompt integration enhances the model's focus on retrieved results through structured attention guidance.

\begin{table*}[!ht]
    \centering
    \begin{tabular}{c|ll|ccc}
    \hline
      \textbf{Dataset} & \multicolumn{2}{c|}{\rule{0pt}{12pt}\textbf{Filtering Candidates}} & \textbf{Words} & \textbf{$I(\tilde{X}; X | Y; Q)$} & \textbf{EM} \\ \hline
      \multirow{5}{*}{NQ} 
        & \multirow{2}{*}{Exact}  & \raggedright Paragraph-Level   & 78.1 & \underline{0.597} & 21.2 \\ 
        &                         & \raggedright Sentence-Level   & 28.4 & 0.561 & 23.8 \\ 
        & \multirow{2}{*}{Greedy} & \raggedright Query \& Answer & 26.2 & 0.562 & 19 \\ 
        &                         & \raggedright Answer          & 18.2 & 0.556 & 24.3 \\ 
        & \multirow{1}{*}{Exact}  & \raggedright Proposition-Level    & 33.6 & \textbf{0.594} & \textbf{27.4} \\ \hline
      \multirow{5}{*}{HotpotQA} 
        & \multirow{2}{*}{Exact}  & \raggedright Paragraph-Level & 120.0 & \textbf{0.679} & 26.3 \\ 
        &                         & \raggedright Sentence-Level & 41.2  & 0.619 & 27.8 \\ 
        & \multirow{2}{*}{Greedy} & \raggedright {\scriptsize Query \& Supporting Facts \& Answer} & 32.5 & 0.614 & 25.8 \\ 
        &                         & \raggedright {\scriptsize Supporting Facts \& Answer} & 14.8 & 0.604 & 26.9 \\ 
        & \multirow{1}{*}{Exact}  & \raggedright Proposition-Level  & 42.6  & \textbf{0.679} & \textbf{28.7} \\ \hline
    \end{tabular}
    \caption{The effectiveness of the information bottleneck theory on the filtering data compression rate and concise mutual information in the test sets of the PGCC module for NQ and HotpotQA. The proposition-level chunking in Hash-RAG, combined with the PGCC module guiding LLMs to focus on the correspondence between propositions and documents, is theoretically optimized and experimentally validated as the optimal trade-off between efficiency and accuracy.}
    \label{tab:qa_performance}
\end{table*}

\section{Analysis}
\subsection{Chunk-Unit \& Prompt-Guided}

\noindent \textbf{Information Bottleneck of Proposition}\ \ \ To demonstrate how HASH-RAG's chunking strategy leverages the information bottleneck to enhance text generation capabilities, we analyze content preserved through document segmentation. Experimental analysis based on Table \ref{tab:qa_performance} reveals a potential correlation between compression rates and the conciseness of conditional mutual information $I(\tilde{X}; X | Y; Q)$, comparing exact and greedy search methods across context lengths for generator, mutual information metrics, and EM scores. We propose that applying information bottleneck principles to factual chunking generates concise intermediate answer representations with supporting evidence, outperforming alternative strategies in multi-hop queries. 
This indicates replacing propositions with complete documents would increased noise, distracted attention, and reduced storage efficiency, thereby degrading generation accuracy. 

\begin{figure}[t]
\centering
\includegraphics[scale=0.41]{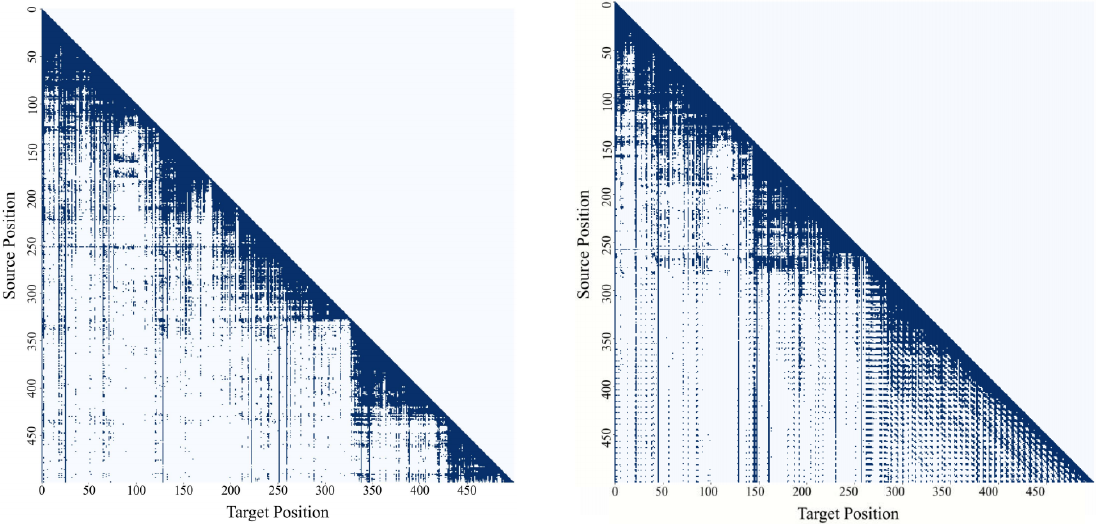}
\caption{Prompt Guidance Attention Heat Map} 
\label{Attention}
\end{figure}
 
\noindent \textbf{Prompt Guidance on Attention}\ \ \ To investigate how prompts influence attention mechanisms during LLM text generation, we employ Recall@1 to identify a document providing optimal factual support, thereby validating the effectiveness of prompt optimization. We generate comparative attention heatmaps (Figure \ref{Attention}) illustrating model behavior with versus without prompts. The prompt-free condition exhibits concentrated self-referential attention along the diagonal axis. In contrast, prompted generation demonstrates vertical attention patterns focusing on proposition tokens $P_j$, accompanied by significant off-diagonal highlights indicating strengthened long-range dependencies between answer generation positions and critical propositions.

\begin{figure}[t]
\centering
\includegraphics[scale=0.3]{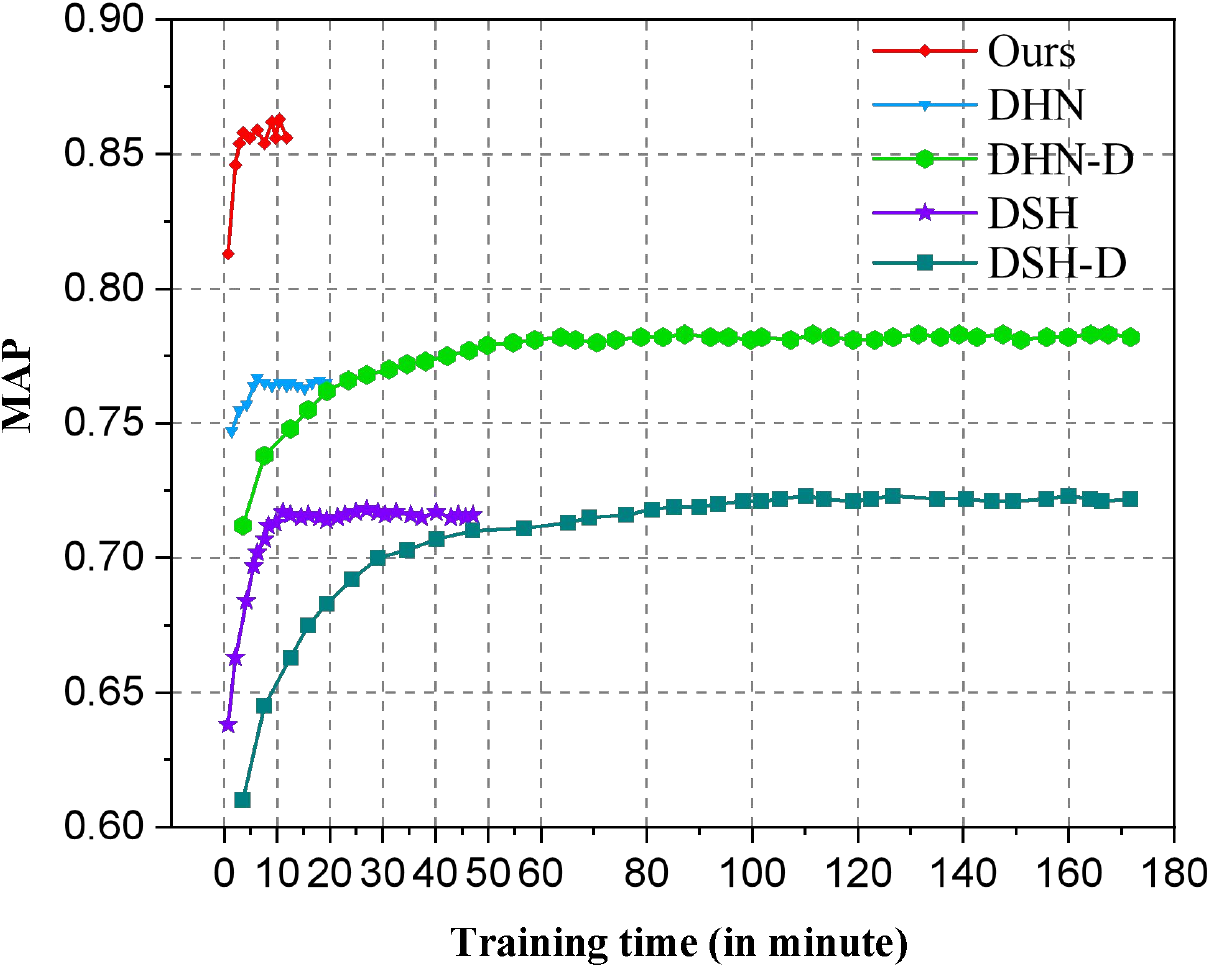}
\caption{Training time on NQ dataset.} 
\label{Ablation_TC}
\end{figure}

\subsection{Training of deep hashing algorithms}

\noindent \textbf{Time Complexity}\ \ \ 
We compare our hash method to other deep hashing baselines on NQ dataset, with the training time results illustrated in Figure \ref{Ablation_TC}. In our evaluation framework, DSH \cite{liu2016deep} and DHN \cite{zhu2016deep} represent conventional deep hashing baselines trained on 10,000 sampled data points, while DSH-D and DHN-D denote their full-database counterparts. The result reveals that full-database training of baselines needs over 80 minutes for convergence, which inspires the sampled training of the large-scale datasets. Moreover, our method achieves significantly faster convergence than both sampled and full-database baselines while maintaining the highest accuracy.

\begin{figure}[t]
\centering
\includegraphics[scale=0.24]{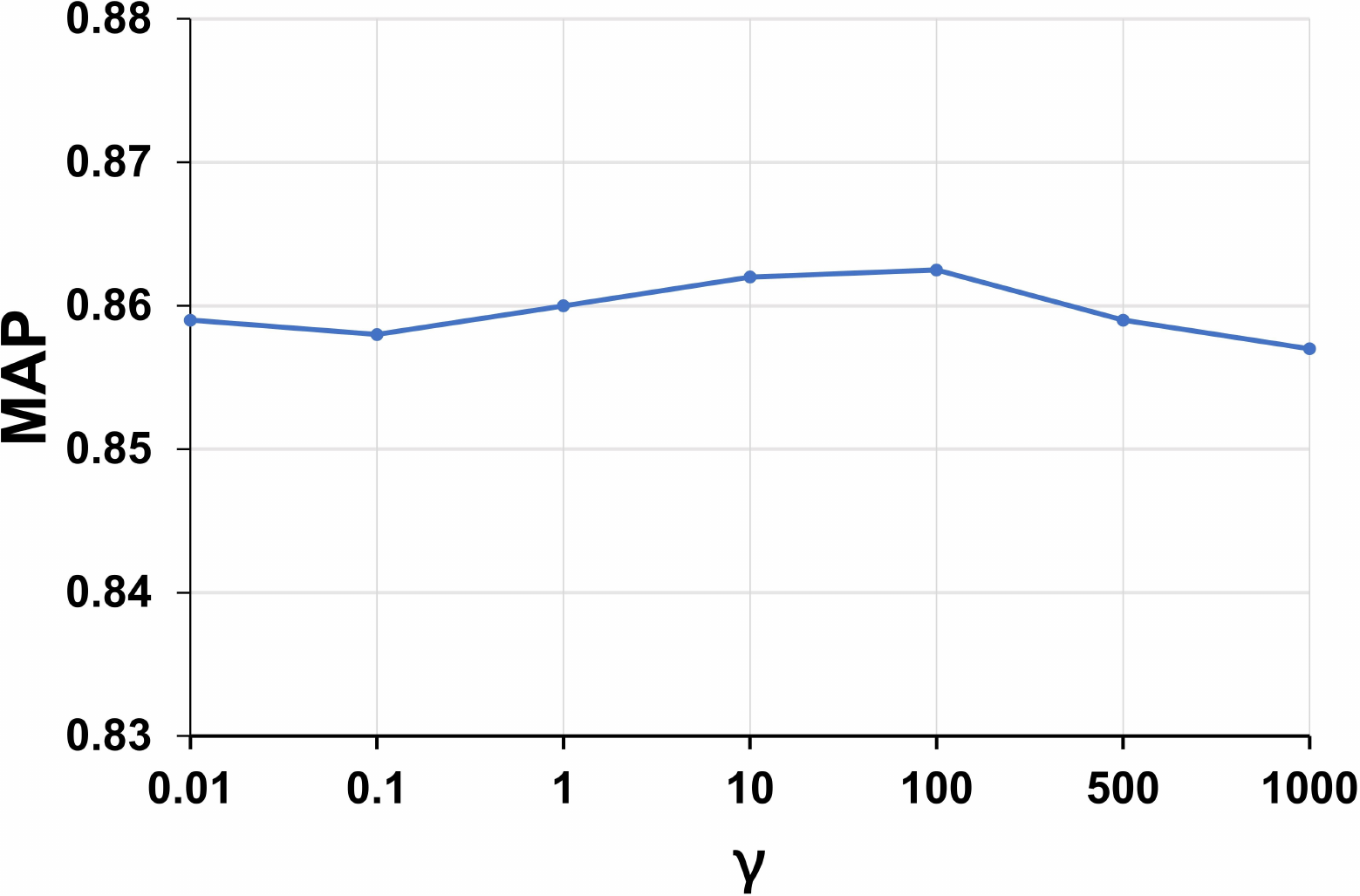}
\caption{Hyperparameter $\gamma$ on NQ dataset.} 
\label{Ablation_S2P}
\end{figure}

\noindent \textbf{Sensitivity to Parameters}\ \ \ Figure \ref{Ablation_S2P} illustrates the hashing hyperparameter $\gamma$ sensitivity on NQ dataset. Our method shows stability across a broad range (1 \textless\ $\gamma$ \textless\ 500), with Mean Average Precision (MAP) fluctuating within 0.01. It could potentially be attributed to NQ's hierarchical semantic structure, which exhibits tolerance to hash-induced local perturbations. This parameter invariance reduces deployment optimization complexity while ensuring multi-scenario reliability.

\section*{Conclusion}
We bridge deep hashing with retrieval-augmented generation for efficient, fine-grained knowledge retrieval and context-augmented generation, balancing the trade-off between the query processing time and recall. Not only as an evaluation framework for hash retrievers, Our proposed PGCC module further improves the accuracy and relevance of retrieval by optimizing the chunking strategy and addressing contextual information limitations. Experimental results demonstrate that our hash retriever significantly outperforms baseline methods and achieves impressive metrics in the generator. In future work, we plan to explore the application of hash techniques to other tasks and structures, such as the knowledge graph.

\section*{Acknowledgments}
This rescarch is supported by the Sichuan International Science and Technology Innovation Coopcration Project with ID 2024YFHZ0317,the Chengdu Scicnce and Technology Burcau Project with ID 2024-YF09-00041-SN,and the National Natural Science Foundation of China Project with ID W2433163.

\section*{Limitations}
The focus of this paper is to deeply integrate deep hashing techniques with the RAG model. The experimental framework assumes that the external knowledge base is static. If incremental updates are required, such as adding new documents or revising content, the hash encoder needs to be retrained to incorporate the new data, which is computationally expensive.  In the future, developing an efficient adaptation strategy for dynamic hashing encoding remains an open challenge.

\bibliography{anthology,custom}
\bibliographystyle{acl_natbib}

\appendix
\definecolor{backgroundgreen}{RGB}{230,242,230} 
\definecolor{bordergreen}{RGB}{150,200,150}     
\definecolor{textcolor}{RGB}{0,0,0}             

\newtcolorbox{qaBox}[1]{
  colback=backgroundgreen,
  colframe=bordergreen,
  title={\normalfont\bfseries #1},
  sharp corners,
  boxrule=1.5pt,
  titlerule=0.8pt,
  fonttitle=\large,
  before upper={\vspace{0.3cm}},
  after upper={\vspace{0.2cm}}
}

\section{Prompt Template}
\label{Prompt Template}
\begin{qaBox}{Open-domain QA for LLaMA-2-7B}

\textbf{Please consider all relevant details in the retrieved segments and offer a concise, informative, and contextually appropriate response. If necessary, carefully review the corresponding indexed documents to support your answer with just a few words:}

\vspace{0.4cm}
\textbf{IdxID:23  Title:} Mount Everest \\
\textbf{Propositions:} Mount Everest is Earth's highest mountain above sea level.

\vspace{0.4cm}
\textbf{IdxID:23  Title:} Mount Everest \\
\textbf{Propositions:} Mount Everest is known in Tibetan as Chomolungma.

\vspace{0.4cm}
\textbf{IdxID:59  Title:} Lhasa Tibetan \\
\textbf{Propositions:} Verbs in Tibetan always come at the end of the clause.

....

\vspace{0.4cm}
\textbf{ID=23  Title:} Mount Everest \\
\textbf{Doc:} Mount Everest, known locally as Sagarmatha or Qomolangma,[note 4] is Earth's highest mountain above sea level, located in the Mahalangur Himal sub-range of the Himalayas. The China–Nepal border 

....



\vspace{0.8cm}
\noindent
\textbf{Question:} What is the highest mountain in the world?\\
\textbf{The answer is:} \underline{Mount Everest}
\end{qaBox}

\section{Dataset}
\label{Dataset}
\subsection{Dataset statistics}
\label{Dataset statistics}

\begin{table}[h]
\centering
\begin{tabular}{lccc}
\toprule
\multirow{2}{*}{\textbf{Dataset}} & \multicolumn{3}{c}{\textbf{\# Examples}} \\
 & \textbf{train} & \textbf{dev} & \textbf{test} \\
\midrule
NQ & 79.2 & 8.7 & 3.6 \\
TQA & 78.8 & 8.8 & 11.3 \\
HOTPOTQA & 88.9 & 5.6 & 5.6 \\
\bottomrule
\end{tabular}
\caption{Dataset statistics}
\label{tab: Dataset statistics}
\end{table}
We use three datasets built from Wikipedia articles as supporting documents for answer, response, and judgment generation, as listed in Table \ref{tab: Dataset statistics}.

\subsection{Units of Wikipedia}
\label{Units of Wikipedia}

\begin{table}[ht]
\centering
\begin{tabular}{lcc}
\toprule
 & \# units & Avg. \# words \\
\midrule
Passages & 41,393,528 & 58.5 \\
Sentences & 114,219,127 & 21.0 \\
Propositions & 261,125,423 & 23.2 \\
\bottomrule
\end{tabular}
\caption{Statistics of text units in the English Wikipedia.}
\label{tab: units Wikipedia}
\end{table}
We refer to the processed corpus as Prop-WIKI. The statistics of Prop-WIKI are shown in Table \ref{tab: units Wikipedia}. Notably, the values presented here correspond to the average segment length of the processed Wikipedia corpus, while Table \ref{tab:qa_performance} specifically reports the average input sequence length fed into the text generator during inference.

\end{document}